# Half-wave plasmonic nanolasers near the localization limit

Sangyeon Cho[1,2]* and Seok-Hyun Yun[1,3]**

[1]Wellman Center for Photomedicine and Harvard Medical School, Massachusetts General Hospital, 65 Landsdowne St., Cambridge, MA 02139, USA

[2]Present address: Department of Bioengineering, Rice University, Houston, TX 77005, USA

[3]Harvard-MIT Health Sciences and Technology, Massachusetts Institute of Technology, Cambridge, MA 02139, USA

Correspondence to: * scho1105@rice.edu or ** syun@hms.harvard.edu

## Abstract

Miniaturized lasers with sub-micron dimensions are of broad interest for optical science, on-chip communication, sensing, and biomedical barcoding. Recently, lowest-order half-wave lasing was demonstrated in semiconductor-on-metal cavities by operating away from the highly dispersive and absorptive surface-plasmon resonance. Here, we demonstrate half-wave-mode lasing near the surface-plasmon resonance at ~630 nm using high-gain indium phosphide (InP) nanoparticles on ultrasmooth gold substrates. The smallest lasing particle, estimated from simulated dispersion curves to have a length of ~115 nm and height of ~100 nm, emitted at 730 nm in air, representing one of the smallest reported active laser cavities. Linewidth and threshold pump fluence generally decreased as the lasing wavelength shifted farther from the plasmon resonance. In larger particles with lengths of 280–480 nm, we observed lasing attributable to second- and third-order plasmonic modes with progressively narrower linewidths. These results extend half-wave dipolar lasing toward near-infrared and visible wavelengths and further push laser miniaturization toward the plasmonic localization limit.

## Introduction

Achieving lasing at the fundamental half-wave condition represents a limiting case of optical cavity miniaturization.[1–3] At this limit, the cavity length approaches approximately one half of the optical wavelength inside the resonator, allowing coherent oscillation in a deeply subwavelength volume while naturally low-order, and potentially single-mode, operation. Such extreme miniaturization is important not only for scaling coherent light sources toward dense photonic integration[4–6], but also for applications in sensing, optical tagging, and biomedical barcoding, where small physical size, spectral distinguishability, and high optical brightness are essential.[7–12]

Plasmonic cavities offer a promising route to this regime because surface plasmon polaritons can confine electromagnetic fields far below the diffraction limit.[13–20] The concept of a deeply subwavelength plasmonic laser was proposed more than two decades ago,[21] and subsequent experiments using metallic nanoparticles and dyes reported signatures of stimulated emission.[22,23] However, unambiguous verification of lasing from individual particles was challenging; the observed emission could also arise from multiparticle coupling, random lasing, or collective effects,[24,25] and the interpretation has not always been fully consistent with conventional laser models.[26,27]

More recently, room-temperature half-wave lasing was demonstrated in semiconductor-on-metal cavities using InGaAsP nanoparticles on metallic substrates, with emission in 1000-1400 nm.[27] In these devices, operation away from the surface-plasmon resonance reduced the severe dispersion and absorption associated with plasmonic polaritons, allowing the lowest-order longitudinal mode to reach threshold. A semiconductor laser model incorporating carrier filling and Fermi blocking successfully explained the observed lasing dynamics, establishing half-wave plasmonic lasing as a physically consistent nanoscale operating principle. Nevertheless, pushing this concept toward shorter wavelengths and closer to the plasmonic resonance remains difficult because metallic absorption increases, plasmonic dispersion becomes stronger, and the cavity quality factor can be severely reduced.[28] These effects are especially important near the surface-plasmon resonance where optical confinement is strong but loss is also substantial. The resonance wavelength is around 630 nm for gold-semiconductor interface.

Here, we investigate whether half-wave plasmonic lasing can be extended toward visible and near-infrared wavelengths while operating close to the surface-plasmon resonance of the gold-semiconductor interface. We use indium phosphide (InP) nanoparticles, with a bandgap wavelength of approximately 920 nm, placed directly on ultrasmooth gold substrates to form deeply subwavelength semiconductor-on-metal cavities. This platform allows us to examine how particle size, plasmonic dispersion, and metal-associated loss shape the transition from the fundamental half-wave mode to higher-order longitudinal plasmonic modes. Through numerical simulations, controlled nanoparticle fabrication, and hyperspectral measurements of individual particles, we analyze resonance evolution, lasing thresholds, linewidths, and mode-dependent Q-factors. This work clarifies both the opportunities and current limitations of half-wave and higher-order plasmonic nanolasing near the plasmonic localization regime.

## Results

### Mode dispersion of semiconductor-on-gold complex: Simulations

We first investigated the fundamental plasmonic half-wave mode of an InP semiconductor particle placed on a semi-infinite gold substrate and surrounded by air (Fig. 1a). The simulated structure consisted of a hexagonal-prism InP particle with a long-axis length L and a height H. The resonance properties of the particle-on-gold structure were examined by calculating Mie scattering spectra using finite-difference time-domain (FDTD) simulations. Figure 1b shows differential scattering cross-section spectra obtained as the particle size was varied. The height was scaled with particle length according to H = 100 nm + 0.35 L. For a given L, the longest-wavelength resonance corresponds to the lowest-order dipolar mode, denoted (1,0) mode, with its dipole moment orientated along the major axis. The next resonance corresponds to a dipolar mode oriented along the minor axis, denoted as the (0,1) mode.

As L decreases, the resonance wavelength of the (1,0) mode initially decreases linearly down to ~ 800 nm. At shorter wavelengths, however, the dispersion becomes increasingly nonlinear and approaches a plateau near the plasmonic dispersion asymptote of the gold/InP interface at ~630 nm. A similar trend is observed for the (0,1) mode. For L < ~180 nm, the particle-gold structure supports only the lowest-order dipolar mode in each polarization direction.

For L > ~200 nm, second-order quadrupolar resonances, denoted as the (1,1) and (2,0) mode, become visible. At L > ~280 nm, the (2,0) mode appears to split into two branches, labeled as $(2,0)_0$ and $(2,0)_1$, presumably due to coupling with a dielectric mode supported within the semiconductor particle. Additional higher-order modes, including (1,2), (2,1), and (0,2), also emerge for L > 200 nm (See Supplementary Fig. 1). While the plasmonic modes exhibit asymptotic cutoffs near the gold/InP resonance wavelength of ~630 nm, dielectric modes supported within InP particles can extend beyond this plasmonic resonance into the yellow-blue spectral range.

The cutoff dimensions and order in which higher-order modes appear depend on particle geometry. For example, Supplementary Fig. 2 shows the dispersion curves for particles with a fixed height of H = 100 nm.l For a particle with L = 125 nm and H = 100 nm, dimensions within our experimental range, Figs. 1c and 1d show the simulated electric-field amplitude, |E|, of the (1,0) mode at λ = 750 nm and the corresponding surface-charge distribution on the gold substrate. The field is tightly confined to the metal–semiconductor interface, as expected for a strongly plasmonic mode. This fundamental mode exhibits a pronounced dipolar surface-charge distribution localized at the semiconductor–gold interface. The associated displacement currents circulate along the particle–metal boundary, producing an in-plane magnetic dipole moment oriented along the minor axis.

To clarify the origin of this mode, we performed scattering simulations for the same particle while varying the gap distance between the InP particle and the gold substrate from 20 nm to 0 nm (Fig. 1e). When the particle is separated from the gold surface, it exhibits a scattering peak at 444 nm, corresponding to a dielectric magnetic-dipole resonance (Supplementary Fig. 3). As the particle approaches the gold surface, this dielectric mode hybridizes with the continuum of surface-plasmon states supported by the gold substrate. This hybridization generates a new interfacial magnetic-dipole resonance that emerges near ~600 nm at a gap distance of 10 nm and progressively red-shifts to ~755 nm when the particle contacts the gold surface. In contrast, the original dielectric resonance remains largely confined within the semiconductor and shifts only slightly. The resulting contact-mode resonance therefore corresponds to a strongly confined plasmonic half-wave mode localized at the metal–semiconductor interface.

FDTD simulations performed for varying L at a fixed height of H = 100 nm further revealed how the electromagnetic energy distribution depends on resonance wavelength (Fig. 1f). As the resonance approaches the plasmonic resonance wavelength of ~630 nm, the modal energy becomes increasingly localized in the gold rather than in the InP particle or surrounding air. By contrast, at longer wavelengths near 1000 nm, ~75% of the modal energy resides in the semiconductor. The simulated quality factor, Q, and peak Purcell factor, $F_p$, of the (1,0) mode reach maximum values of approximately 16 and 270, respectively, around 700 nm. This wavelength range coincides with the minimum optical absorption of gold near ~720 nm (Supplementary Fig. 4). The simulated Q factor was largely insensitive to particle shape, showing similar values for cylindrical and cuboidal geometries (Supplementary Fig. 5).

**Half-wave mode lasers: Experiments**

InP nanolaser particles were fabricated using a combination of dry and wet etching processes. Briefly, maskless optical lithography, followed by oxygen plasma trimming and reactive-ion etching, was used to define the initial particle geometry. Subsequent wet etching in piranha solution released the particles and isotropically reduced their dimensions. Figure 2a shows SEM images of representative InP particles drop-cast onto gold substrates. The long-axis lengths of ten randomly selected particles ranged from 120 to 210 nm, with a mean value of 163 nm (Supplementary Fig. 6).

Atomic force microscopy measured a roughness of 1.2 nm of the gold surface. SEM imaging revealed polycrystalline domains on the gold surface with lateral sizes of approximately 150–450 nm (Fig. 2a; Supplementary Fig. 7). Given the particle dimensions, an individual InP particle is expected to contact one or two gold crystalline domains. Adding a ~ 5 nm silica spacer layer on the gold surface did not noticeably affect the lasing performance, presumably because the intrinsic surface roughness of the InP particles already creates a nanoscale separation between the semiconductor and the gold surface. The particles were optically pumped using a 532 nm nanosecond laser with a repetition rate of 10 kHz and a pulse width of 4 ns. The output emission was analyzed using a grating-based spectrometer with a spectral resolution of ~0.7 nm (Supplementary Fig. 8).

Figure 2b shows representative emission spectra from a particle lasing at 788 nm as the pump fluence was increased from 0.2 to 8.7 times the threshold fluence, with $P_{th}$ = 1.1 mJ $cm^{-2}$. This gives a pump energy of ~ 0.2 pJ projected over the particle, but the actual absorbed pump energy would be moderately different from this geometry-based estimate due to cavity resonance and reflection from gold. With increasing pump fluence, the broadband fluorescence background clamps, while a narrow emission peak emerges and blue-shifts, consistent with band-filling behavior and characteristic lasing dynamics. Based on the tuning curve in Supplementary Fig. 2a, the observed lasing wavelength corresponds to an estimated particle length of L ≈ 135 nm. The experimental spectral evolution was well reproduced by numerical simulations using the waterfall semiconductor laser model developed in our previous work,[27] with a cavity quality factor of Q = 15 and Purcell factor of F = 8 (Fig. 2c). The corresponding light-in–light-out curves exhibited nonlinear threshold behavior (Supplementary Fig. 9). At the highest pump fluence, the stimulated-to-spontaneous emission ratio reached ~30.

The spectral linewidth was typically narrowest just above threshold and increased modestly with increasing pump fluence, likely because of gain-induced refractive-index fluctuations and carrier-density noise. At pump fluences above ~4 mJ $cm^{-2}$, the output power began to roll off due to Auger recombination limiting the carrier population at high excitation densities. Additional examples of particles lasing at 758 nm (L ≈ 122 nm) and 834 nm (L ≈ 147 nm) are shown in Supplementary Fig. 9.

Figure 2e summarizes emission spectra collected from 35 particles on a gold substrate at pump fluences approximately two to four times above threshold. The spectra are arranged according to particle length estimated by comparing with the simulated tuning curves for the (1,0) and (0,1) modes. The smallest lasing particle had an estimated length of L ≈ 115 nm and emitted at 732 nm. Particles with lengths between ~140 and 165 nm often showed simultaneous lasing from the (1,0) and (0,1) modes, consistent with the spectral overlap of the two dipolar branches in this size range. Larger particles exhibited single-mode lasing corresponding to the (0,1) mode.

We next compared the measured lasing linewidths from the 35 InP particles with previously reported data from two batches of InGaAsP particles with different alloy compositions and dimensions (Fig. 3). The InP data are shown as cyan circles, while the previously reported InGaAsP data are shown as light and dark green circles. Across the combined dataset, the linewidths decrease from ~20–30 nm at higher photon energies of ~1.6–1.7 eV, reach a minimum of ~6–8 nm near ~1.2–1.4 eV, and increase again toward lower photon energies. Numerical simulations with Q = 12 and Q = 15 at a mode-volume-averaged Purcell factor F = 8 capture the overall magnitude and spectral trend, with higher Q values producing systematically narrower linewidths.

The observed nonmonotonic linewidth trend likely reflects several coupled factors, including the wavelength dependence of the plasmonic cavity Q factor, differences in semiconductor gain among the material systems, and the ratio of pump photon energy to bandgap energy. Device-to-device variations in particle geometry, surface quality, and material inhomogeneity likely account for much of the experimental scatter. The sample group near 1.2 eV also had larger semiconductor particle heights than the other groups, which increased the modal Q factor and contributed to the narrower observed linewidths. For identical particle shapes, half-wave nanolasers are therefore expected to exhibit a more modest linewidth minimum in a spectral range of 900–950 nm.

**Higher-order mode lasing**

We next investigated lasing from higher-order resonances in InP particles with lateral sizes ranging from 280 to 480 nm on gold substrates. Figure 4a shows representative SEM images of particles from a fabrication batch with a mean diameter of 335 nm. Because the wet-etch duration was relatively short for these batches, the particles largely retained a disk-like geometry with heights of 300–400 nm, substantially larger than the H ≈ 100 nm used for the half-wave devices. FDTD simulations indicated that these particles support the second-order (2,0) mode in the spectral range of 750–850 nm, with a cold-cavity Q factor of ~19 and a Purcell factor of ~17.

Figure 4b shows measured emission spectra from eight particles, arranged along the calculated tuning curve for the second-order (2,0) mode. In addition to the narrow lasing peaks, broader spectral envelopes are also present, which are likely associated with adjacent cavity modes having lower gain and/or higher loss. Figure 4c shows representative spectra from one particle measured at pump fluences below and above the lasing threshold of 1.2 mJ $cm^{-2}$. At the highest pump level, the laser exhibited a linewidth of ~6 nm. The simulated near-field intensity distribution of the (2,0) mode (Fig. 4d) shows a four-lobe standing-wave pattern along the major axis. The measured far-field emission profile is in good agreement with the simulation (Fig. 4e).

We next examined particles from a second fabrication batch with a mean diameter of 427 nm and heights of 300–400 nm. Figure 5a shows representative SEM images of these particles. Figure 5b presents emission spectra from 15 different particles, arranged along the calculated tuning curve for the (1,2) mode (m = 3). FDTD simulations indicated a cold-cavity modal Q factor of ~30 and a Purcell factor of ~6 for this mode. Figure 5c shows representative spectra from an individual particle measured near the lasing threshold of 0.9 mJ $cm^{-2}$. The third-order mode laser exhibited a substantially narrower linewidth of ~1.8 nm. The ratio of the center wavelength to the linewidth, corresponding to the lasing Q factor, was ~450, approximately 15 times greater than the cold-cavity modal Q factor.

The simulated near-field intensity distribution of the third-order mode (Fig. 5d) exhibits a distinct multi-lobe standing-wave pattern along the particle axes, consistent with the expected charge distribution of an m = 3 plasmonic resonance. The experimentally measured far-field emission profile, which shows three pronounced intensity lobes, is also in good agreement with the simulation. Although the long-axis length distributions of the m = 2 and m = 3 batches partially overlap, batch-dependent differences in height and etch-induced ellipticity likely led to different mode dispersion and lasing modes.

**Modal scaling of lasing linewidths and threshold pump fluence**

Figure 6a summarizes the lasing quality factors of the plasmonic nanolasers for the three modal families. The extracted lasing Q factors show a clear increase with modal order. The fundamental half-wave devices, corresponding to m = 1, exhibit lasing Q values of approximately 30–90. In contrast, the second- and third-order modes reach substantially higher values of approximately 120–250 and 350–600, respectively. Linear fits further indicate that the wavelength dependence of the lasing Q factor becomes stronger with increasing modal order, with slopes of approximately 0.2, 0.56, and 2.1 $nm^{-1}$ for m = 1, 2, and 3, respectively. Higher-order modes benefit more strongly from reduced plasmonic loss at longer wavelengths, producing narrower lasing linewidths and higher spectral coherence.

The threshold pump fluence also decreased substantially with increasing mode order (Fig. 6b). For each modal family, the threshold generally decreased at longer lasing wavelengths, although significant particle-to-particle variation was observed. This variation likely reflects differences in particle geometry, surface quality, and coupling to adjacent modes. Many half-wave nanolasers and nearly all third-order devices exhibited thresholds below 10 pJ $\mu m^{-2}$, despite their deeply subwavelength dimensions. These values are comparable to those of much larger semiconductor-only laser particles with diameters of approximately 2 µm operating in the 1100–1600 nm spectral range. Together, these results indicate that higher-order plasmonic modes can provide improved linewidth and threshold performance, although the trade-off is an increase in cavity size relative to the fundamental half-wave devices.

## Discussion

We investigated InP-on-gold nanolasers operating in the 710–870 nm spectral range, close to the plasmonic dispersion asymptote of the gold–semiconductor interface. Lasing was observed from InP particles with major-axis lengths as small as ~115 nm. These dimensions are smaller than those recently reported for InGaAsP particles lasing at longer wavelengths of 1000–1400 nm. The size reduction is greater than expected from simple linear wavelength scaling because of the nonlinear plasmonic dispersion that occurs near the surface-plasmon resonance. The effective mode volume approached approximately $(\lambda/4n)^3$, where n=3.5 is the refractive index of the semiconductor. This volume is approximately one-eighth of the conventional half-wave cavity volume, $(\lambda/2n)^3$, achievable using a semiconductor cavity alone.

Because the plasmonic dispersion asymptote depends on the surrounding refractive index and the optical properties of the metal, the cutoff wavelength can in principle be tuned by selecting appropriate metal-semiconductor combinations. In particular, the cutoff could be shifted further into the visible spectral range using gain media with lower refractive indices and suitable bandgaps, such as CdS, or CdSe on gold. Such material combinations may allow half-wave plasmonic lasing across a broader portion of the visible spectrum while maintaining deeply subwavelength mode volumes.

Beyond optical pumping, electrical injection of these deeply subwavelength cavities represents an important step toward integrated coherent light sources.[29] Half-wave metal–semiconductor nanolasers may therefore provide a platform for coherent emitters operating near the ultimate confinement limits of plasmonic photonics.[30,31] At the same time, the present results also highlight remaining challenges, including metal-associated absorption, surface roughness, contact variability, and carrier-loss mechanisms such as Auger recombination at high pump densities.

The substrate-based geometry used here provides a convenient platform for studying individual semiconductor-on-metal cavities. However, techniques also exist for depositing or coating metals directly onto semiconductor particles with appropriate insulating spacer layers.[27,32,33] With optimized metal coating, it may be possible to realize free-standing nanolaser particles operating at the lowest-order half-wave plasmonic mode. Such particles would behave optically like diffraction-limited dipolar emitters, analogous in spatial emission scale to single molecules or quantum emitters, but with much brighter, narrowband, and coherent emission. This direction could enable compact laser particles for multiplexed optical tagging, sensing, and nanoscale photonic integration.

## Methods

**Device fabrication.** InP particles were fabricated from MOCVD-grown epitaxial wafers containing an InP gain layer on an InGaAsP sacrificial layer atop an InP substrate. For particles targeting the fundamental half-wave mode ($m$ = 1), the InP layer thickness was 270 nm; for higher-order modes ($m$ = 2–3), the InP layer thickness was 500 nm. The InGaAsP sacrificial layer was approximately 250 nm thick in both cases. An ~800 nm-thick SU-8 6000.5 photoresist layer was spin-coated at 1000 rpm and patterned using a maskless aligner with an exposure dose of 3500 mJ $cm^{-2}$. After post-exposure bake, development, IPA rinsing, and hard baking at 180 °C, the SU-8 patterns were trimmed by oxygen plasma ashing, reducing the feature diameter from ~840 nm to ~300–340 nm. The trimmed patterns were transferred into the semiconductor layers by reactive-ion etching at 180 °C using a standard III–V etching recipe, producing nanopillars that fully penetrated the InP gain layer while leaving part of the InGaAsP sacrificial layer intact. The InGaAsP sacrificial layer was selectively removed by wet etching in piranha solution (1:1:10, $H_2SO_4$:$H_2O_2$:$H_2O$) at room temperature for ~20 min, releasing the InP particles. Extended wet etching was used when needed to isotropically reduce the particle dimensions at a rate of ~4 nm $min^{-1}$. After reaching the target size, the etching was quenched, and the particles were rinsed with ethanol and transferred onto ultrasmooth Au substrates (Platypus Technologies) by drop-casting.

The final particle sizes were selected to match the simulated resonance regimes. Particles targeting the fundamental half-wave mode had a mean long-axis length of 163.3 nm after wet etching. Particles targeting higher-order resonances had mean lateral sizes of 335.4 nm for $m$ = 2 modes and 427.2 nm for $m$ = 3 modes. Because the higher-order particles underwent shorter wet etching, they largely retained a disk-like geometry. In contrast, the smaller $m$ = 1 particles underwent extended etching, leading to a hexagonal-prism morphology due to crystallographic anisotropy in the InP etch rate.

**Finite-difference time-domain simulations.** Finite-difference time-domain (FDTD) simulations were performed using Lumerical FDTD Solutions. The optical constants of gold were obtained from prior ellipsometry measurements, and the real part of the refractive index of InP was set to 3.5. Scattering spectra were calculated using a total-field scattered-field source with plane-wave excitation. The scattered power was collected by a power monitor enclosing the structure, and the incident polarization and propagation direction were varied to identify TE- and TM-like resonances. Cavity modes were further analyzed using electric or magnetic dipole sources embedded in the semiconductor particles. The fundamental plasmonic mode was most efficiently excited by an electric dipole oriented parallel to the metal substrate. Time-domain electric and magnetic fields were recorded using point monitors placed in and around the particles, and the spectral response was obtained by Fourier transforming the time-domain signals. Polarization charge distributions were calculated from the divergence of the electric field, and far-field emission patterns were obtained using a box monitor enclosing the metal–semiconductor structure. Perfectly matched layers were used as boundary conditions. Mesh refinement at metal interfaces used the conformal variant 1 scheme, with a mesh size below 5 nm.

The effective mode volume was calculated from the frequency-domain electric-field distribution as $V_{eff} = \int \varepsilon(r)|E(r)|^2 dV \,/ max_{r \in InP}[(r)|E(r)|^2]$ where the denominator was evaluated within the semiconductor region only. For dispersive materials such as gold, the generalized electromagnetic energy density incorporating frequency-dependent permittivity was used. Material-specific energy fractions were calculated by integrating the corresponding energy density over Au, InP, and air regions identified from refractive-index masks. The fractional energy in each material was defined as $\eta_v = U_v / \sum U_v$, where $U_v$ is the integrated modal energy in material region $v$.

**Optical experiments.** Optical measurements were performed using a home-built hyperspectral microscope. The pump source was a frequency-doubled Nd:YAG laser operating at 532 nm, with a repetition rate of 10 kHz and a pulse duration of 4 ns. A 50x air objective lens (Nikon, NA = 0.6) was used for both excitation and collection. The emitted light was split into two detection paths: one directed to a silicon-based EMCCD camera (Luca, Andor) for wide-field imaging, and the other directed to a spectrometer (Shamrock, Andor) equipped with an EMCCD detector. Two interchangeable gratings were used: a 300 lines $mm^{-1}$ grating with a 500 nm blaze and a 100 μm slit, providing a spectral resolution of 0.7–0.9 nm; and a 1200 lines $mm^{1}$ grating with a 500 nm blaze and a 100 μm slit, providing a spectral resolution of 0.13 nm.

## Acknowledgements

We acknowledged funding from the National Institutes of Health (R01-EB033155, R01-EB034687). S.C. acknowledged support from the National Institutes of Health (K25CA301033). This research used the resources of the Center for Nanoscale Systems, part of Harvard University, a member of the National Nanotechnology Coordinated Infrastructure, supported by the National Science Foundation under award number 1541959.

## Conflict of interest

S.H.Y. have financial interests in LASE Innovation Inc., a company focused on commercializing technologies based on laser particles. The financial interests of S.H.Y. were reviewed and are managed by Mass General Brigham in accordance with their conflict-of-interest policies. S.C. declare no conflict of interest.

## Author contributions

S.C. and S.H.Y. designed the study. S.C. performed experiments and FDTD simulation. S.C. and S.H.Y. analyzed the data and wrote the manuscript.

## Data availability

The laser experiment data used within this paper are available through. The authors can provide additional data that supports the findings of this study upon a reasonable request.

**Code availability**

The details of FDTD simulations and semiconductor laser modeling can be found in Methods and Supplementary Information. The associated codes can be obtained from the authors upon a reasonable request.

## Figures and Captions

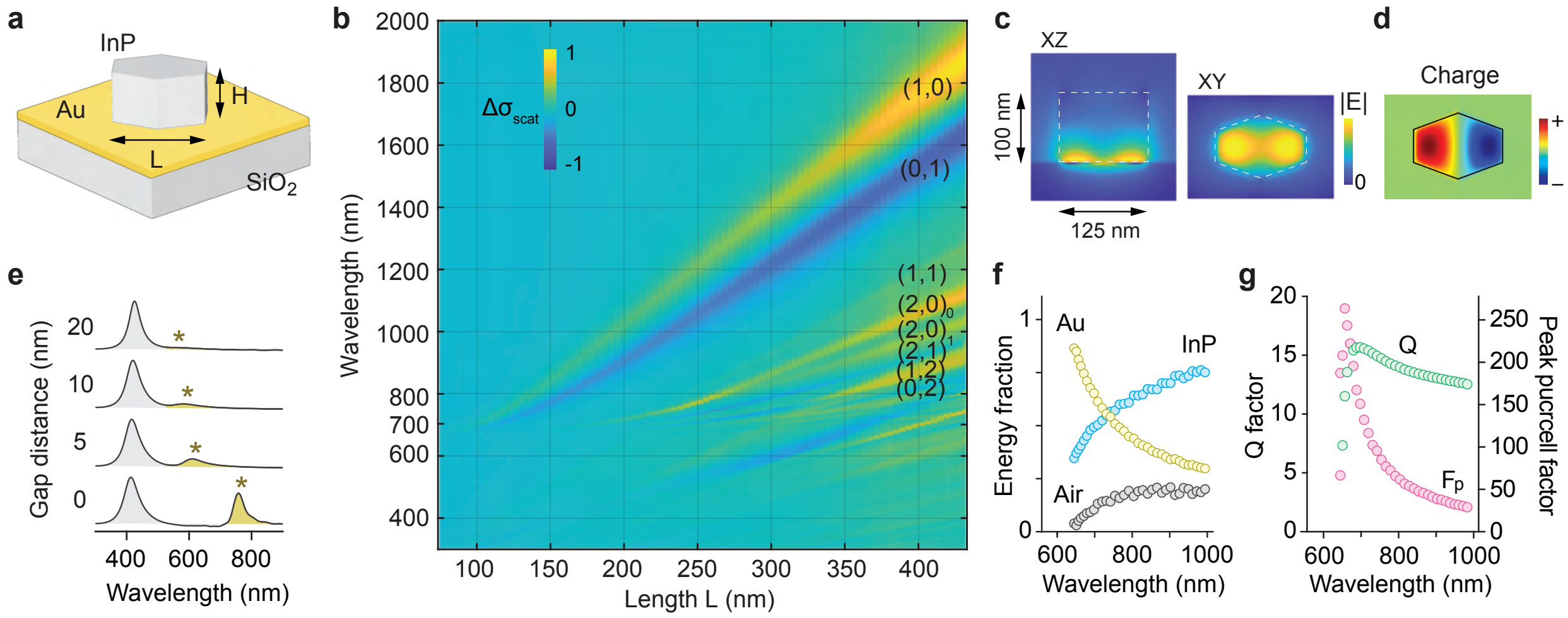

**Fig. 1. Plasmonic half-wave resonance. a**, Schematic of an InP-on-gold consisting of a hexagonal InP particle on a laterally semi-infinite gold (Au) with 100 nm thickness on $SiO_2$ substrate**. b**, Differential scattering cross-section map, $\Delta\sigma_{scat} = \sigma_x - 1.3\ \sigma_y + \sigma_z$, plotted as a function of particle length *L*. Here, $\sigma_x$ and $\sigma_y$ denote the scattering amplitudes under top-incident excitation with the incident magnetic field polarized along the particle minor and major axes, respectively. $\sigma_z$ denotes the scattering amplitude under side-incident excitation with the incident magnetic field polarized out of plane, for which the associated electric field is oriented along the particle minor axis. The particle height (H) was scaled proportionally according to H = 100 nm + 0.35 L. **c**, Simulated electric-field amplitude |E|, in the XZ vertical and XY horizontal planes for the lowest-order (1,0) mode at 750 nm. **d**, Simulated dipolar charge distribution of the (1,0) mode at the gold surface. **e**, Mie-scattering spectra as a function of particle-metal gap distance, showing coupling between the dielectric magnetic dipole resonance of the particle and the surface-plasmon mode. Asterisks indicate the gap plasmonic mode. **f**, Wavelength-dependent fractional modal energy stored in Au, InP, and air for the (1,0) mode. **g**, Wavelength-dependence of the cavity quality factor, Q, and peak Purcell factor, $F_p$, for the (1,0) mode. In the simulations in c-g, L = 125 nm and H = 100 nm.

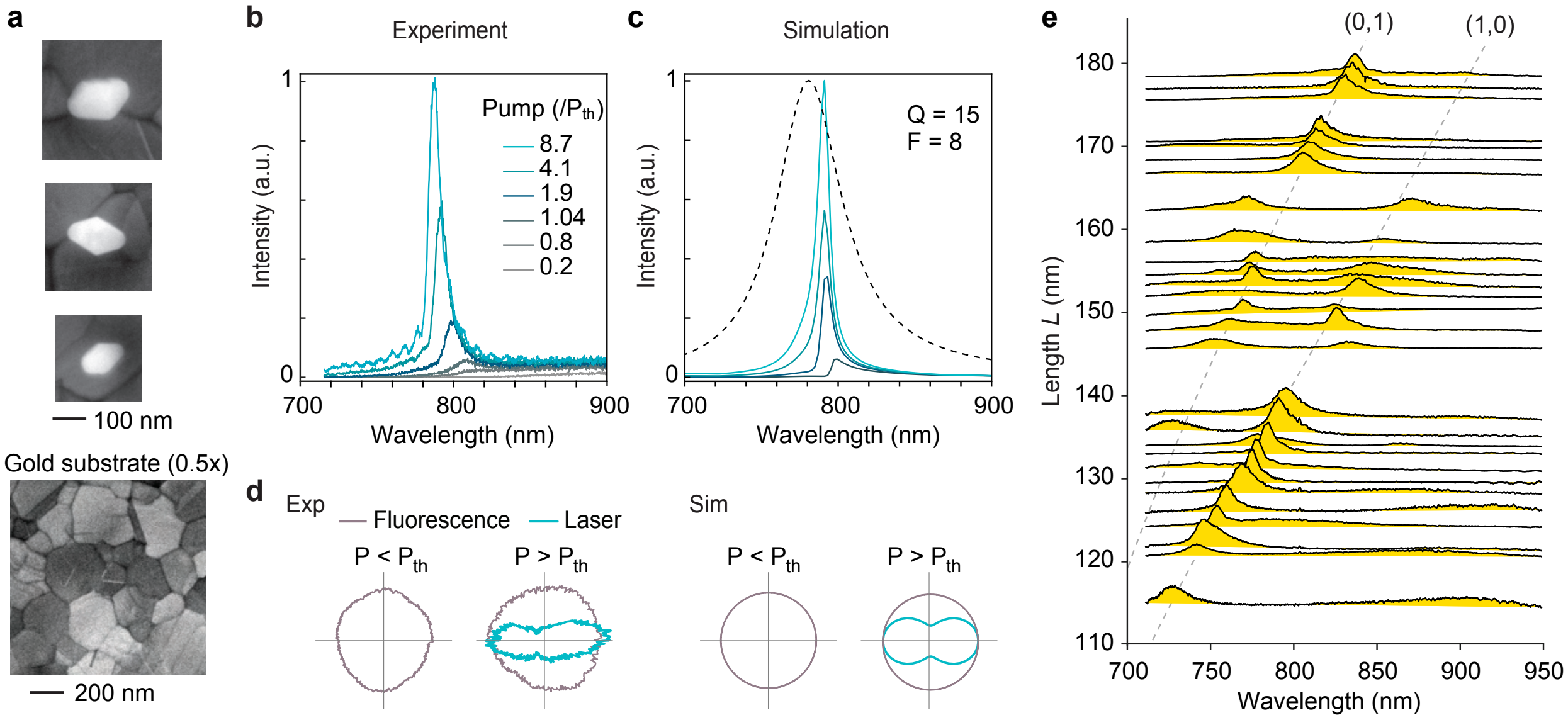

**Fig. 2. Lasing characteristics of lowest-order half-wave plasmonic devices. a**, SEM images of three representative InP particles, top, and a typical gold substrate, bottom, shown at half the magnification of the particle images. The gold polycrystalline domains are approximately 200–500 nm in size, larger than the InP particles. **b**, Representative emission spectra from an InP-on-gold sample measured at different pump powers. **c**, Simulated emission spectra for Q = 15 and a spatially-averaged Purcell factor F = 8. **d**, Measured output polarization states from a half-wave laser below and above the lasing threshold, left, showing good agreement with simulations, right. **e**, Emission spectra from 35 particles measured above threshold and arranged along the dispersion curves of the lowest-order (1,0) and (0,1) modes.

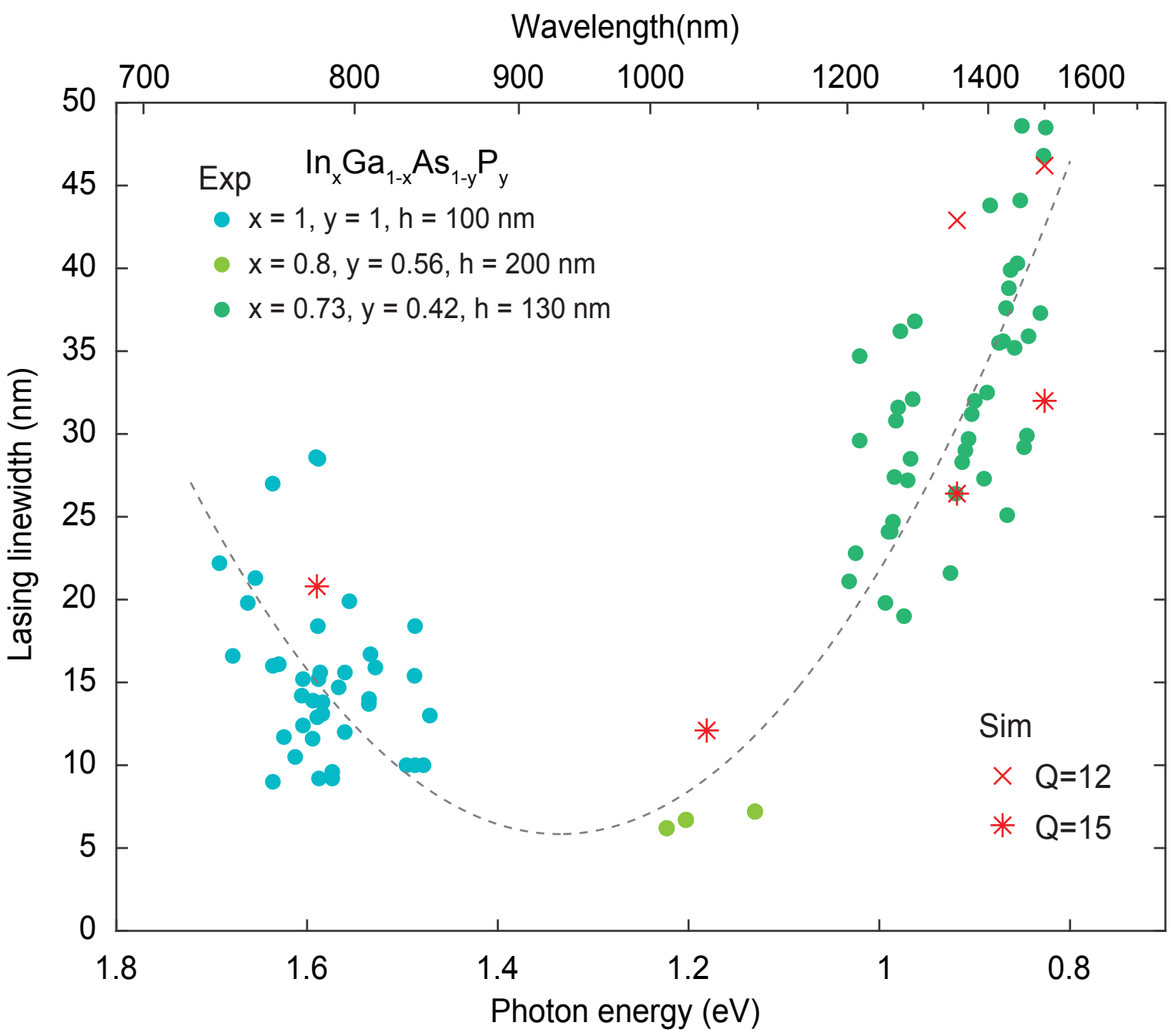

**Fig. 3. Laser linewidth comparison.** Lasing linewidths of the InP-on-gold devices reported in this work are shown as blue circles. Previously reported devices in Ref. 27 are shown as light and dark green circles. Semiconductor compositions (x, y) and particle heights (h) are indicated. Crosses (×) and asterisks (*) indicate simulation results for estimated particle dimensions with cavity Q-factor of 12 or 15 and a spatially averaged Purcell factor of 8. The dashed curve is provided as a visual guide. The sample group near 1.2 eV had larger particle heights than the other groups, which contributed to their lower linewidths.

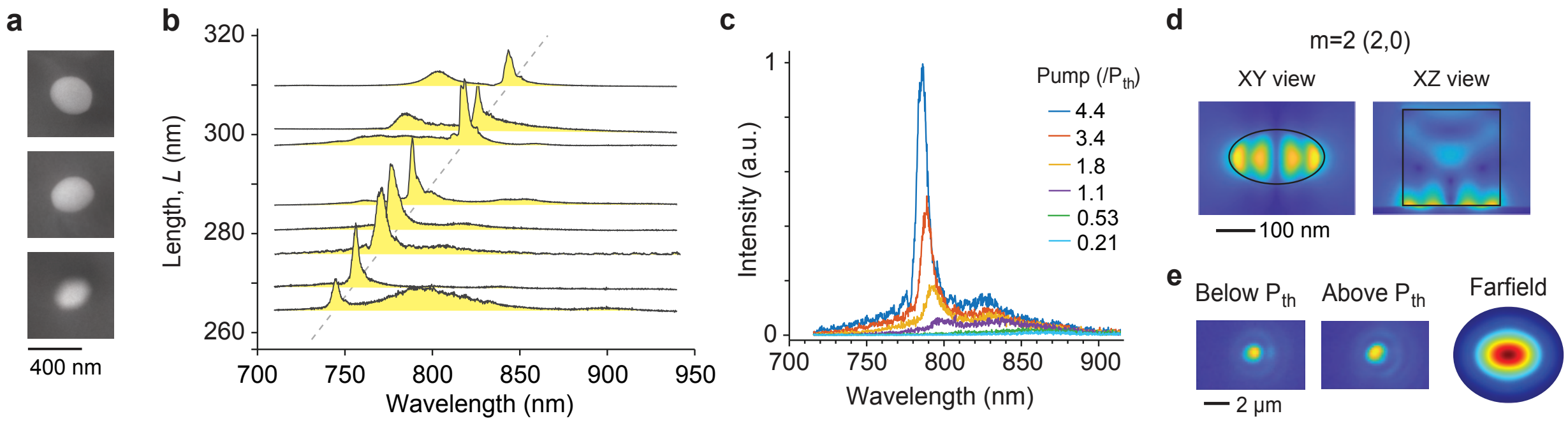

**Fig. 4. Lasing of second-order plasmonic modes. a**, SEM images of particles in a second fabrication batch, with diameters ranging from 260 to 320 nm. **b**, Measured emission spectra, arranged along the dispersion curve of the (2,0) mode, shown as a dashed line. **c**, Representative pump-dependent emission spectra from a single particle. **d**, Simulated near-field amplitude of the (2,0) mode. **e**, Experimental far-field emission patterns below and above threshold, left, and simulated far-field pattern, right.

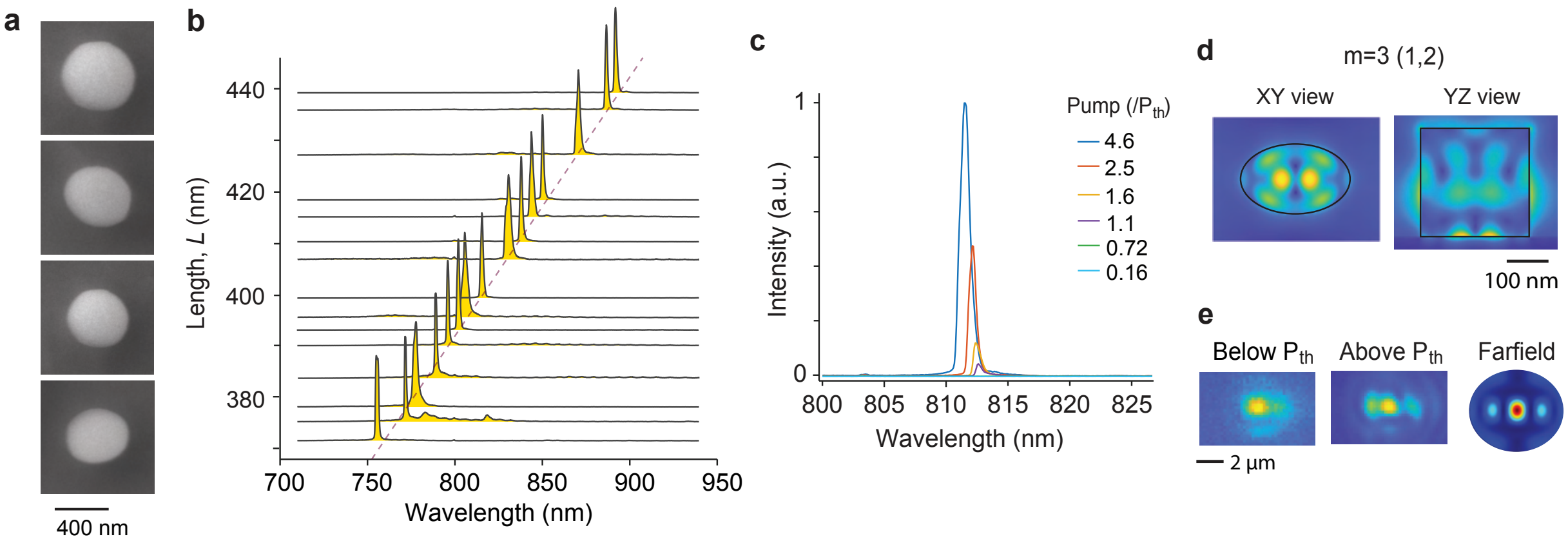

**Fig. 5. Lasing of third-order plasmonic modes. a**, SEM images of particles in the third fabrication batch, with diameters ranging from 370 to 440 nm. **b**, Measured emission spectra arranged along the dispersion curve of the (2,1) mode, shown as a dashed line. **c**, Representative pump-dependent emission spectra from a single particle. **d**, Simulated near-field amplitude of the (1,2) mode. **e**, Experimental far-field emission patterns below and above threshold, left, and simulated far-field pattern, right.

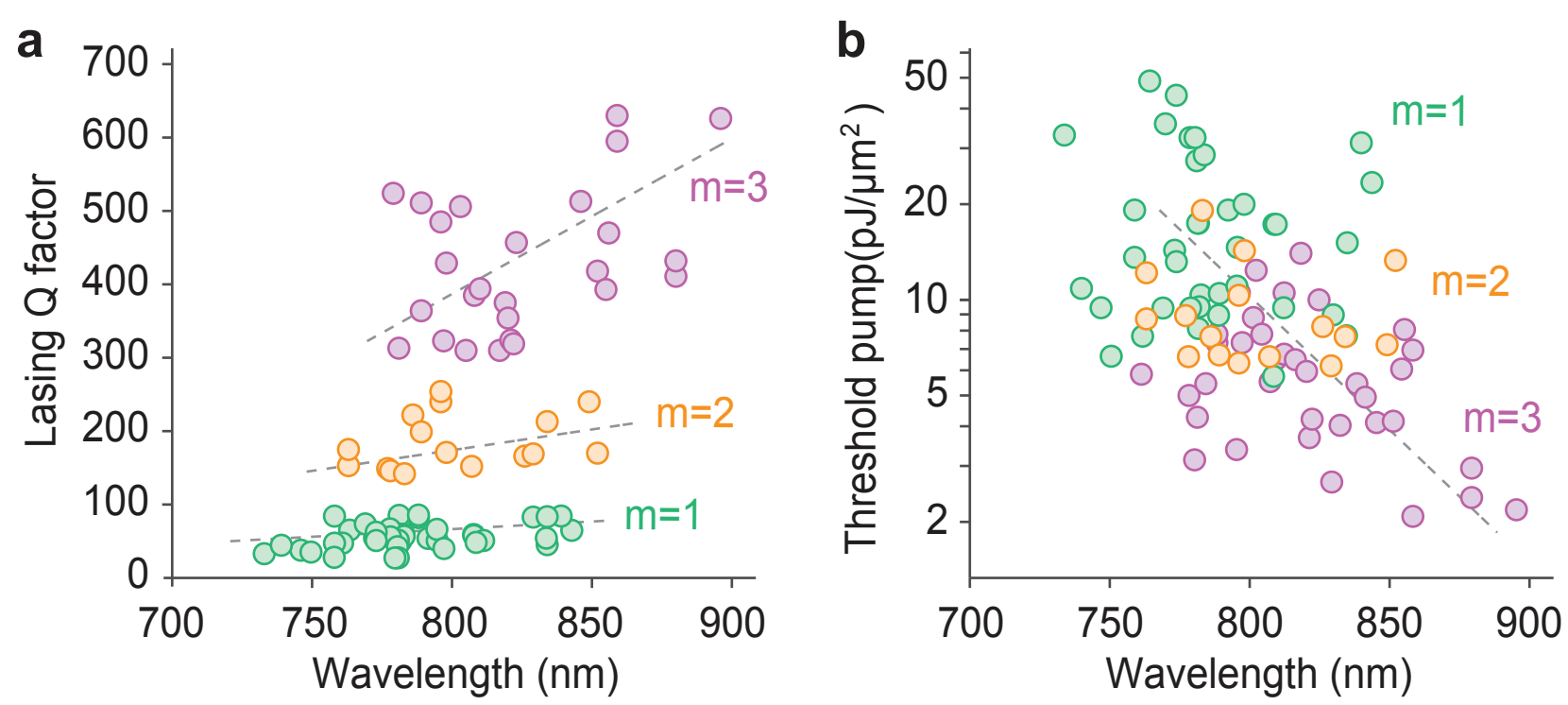

**Fig. 6. Modal scaling of lasing characteristics. a**, Lasing Q factors of devices with different sizes for the three modal families, m = 1 to 3. **b,** Threshold pump fluences of devices lasing in different modal families. Dashed lines indicate linear regressions.

Supplementary Information for

# Half-wave plasmonic nanolasers near the localization limit

Sangyeon Cho[1,2]* and Seok-Hyun Yun[1,3]**

[1]Wellman Center for Photomedicine and Harvard Medical School, Massachusetts General Hospital, 65 Landsdowne St., Cambridge, MA 02139, USA

[2] Present address: Department of Bioengineering, Rice University, Houston, TX 77005, USA

[3]Harvard-MIT Health Sciences and Technology, Cambridge, MA 02139, USA

Correspondence to: *scho1105@rice.edu or **syun@hms.harvard.edu

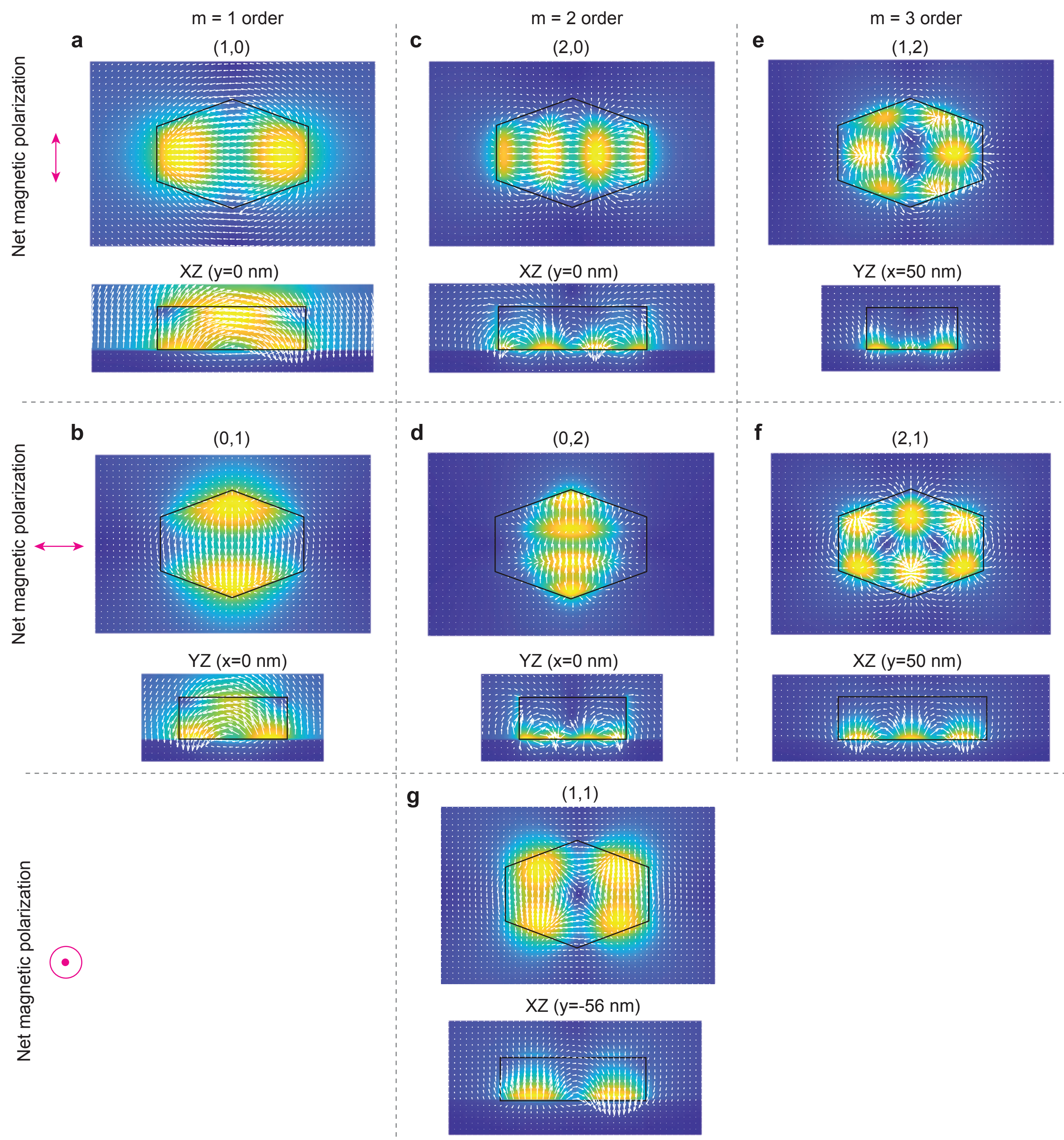

**Supplementary Fig. 1. Electric-field distributions of resonant plasmonic modes. a–g**, Electric-field vector and intensity distributions of representative low-order resonant plasmonic modes in three modal families, m=1–3. The magenta symbols on the left indicate the direction of the incident magnetic-field polarization used to selectively excite each mode in the TFSF-source Mie scattering simulations. The (1,0), (2,0), and (1,2) modes are excited by magnetic-field polarization along the particle major axis, whereas the (0,1), (0,2), and (2,1) modes are excited by magnetic-field polarization along the minor axis. The (1,1) mode is excited by an out-of-plane magnetic-field polarization along the z-axis, denoted by the ⊙ symbol.

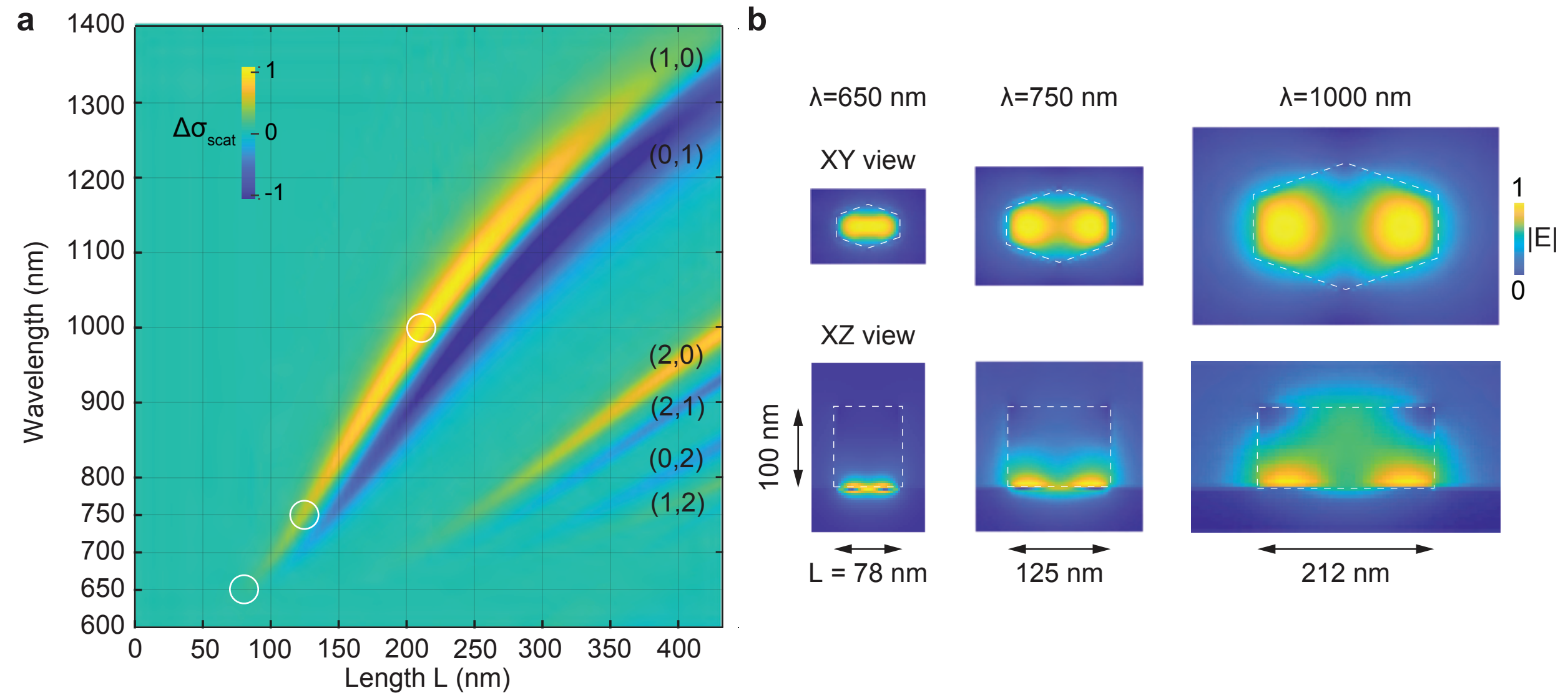

**Supplementary Fig. 2. Simulated modes in particles with varying length and fixed height. a,** Calculated mode dispersion as a function of particle length L for particles with a fixed height of 100 nm. The fixed height constraints the vertical, z-axis, field profile, leading to bending of the (1,0) and (0,1) dispersion branches for L $\gtrsim$ 250 nm. The dispersion branches were obtained from TFSF-source Mie scattering simulations with the incident magnetic-field polarization oriented along the particle major or minor axis. **b,** Simulated electric-field amplitude, |E|, of the fundamental (1,0) mode at λ = 650, 750, and 1000 nm. These modes are indicated by circles in **a**. The XY-plane views show two electric field lobes at the gold-semiconductor interface, while the XZ cross-sectional views reveal strong vertical electric field localized at the interface.

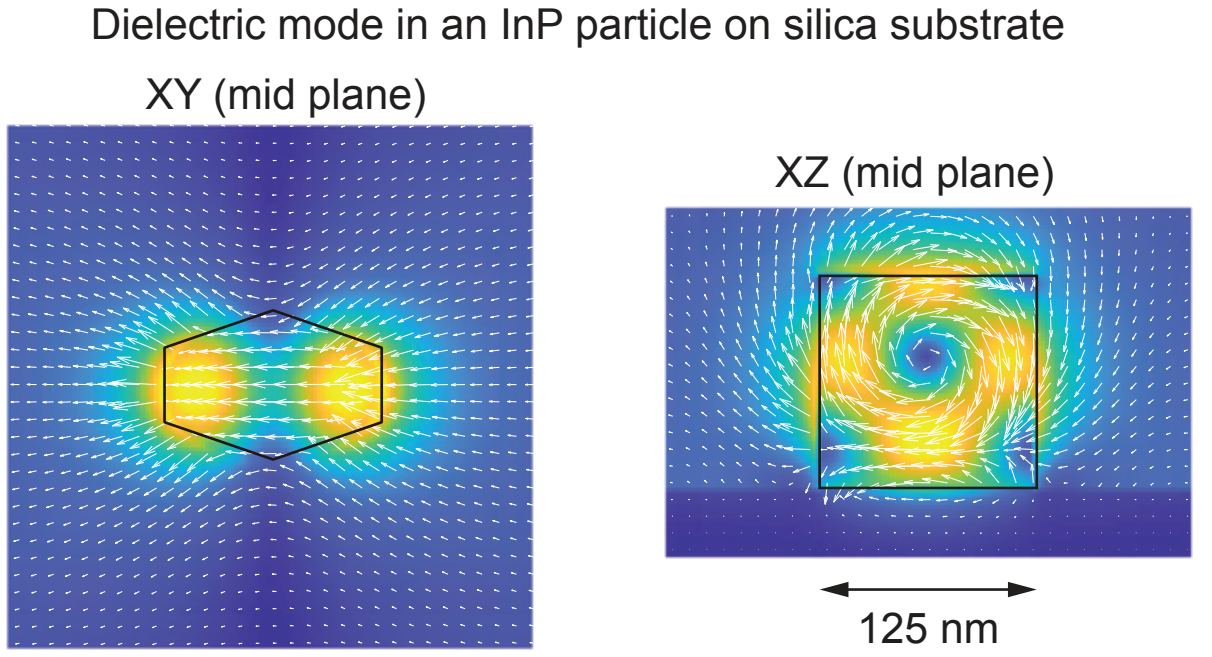

**Supplementary Fig. 3. Magnetic-dipole mode in a dielectric III–V particle. a,** Schematic of circulating displacement currents generating an in-plane magnetic moment inside the dielectric particle. **b,** Simulated electric-field vector and intensity distributions of the fundamental dielectric magnetic-dipole mode in XY and XZ views. Colors indicate normalized field magnitude, and arrows denote field vectors. This mode has a Q factor of ~6 and a Purcell factor near unity.

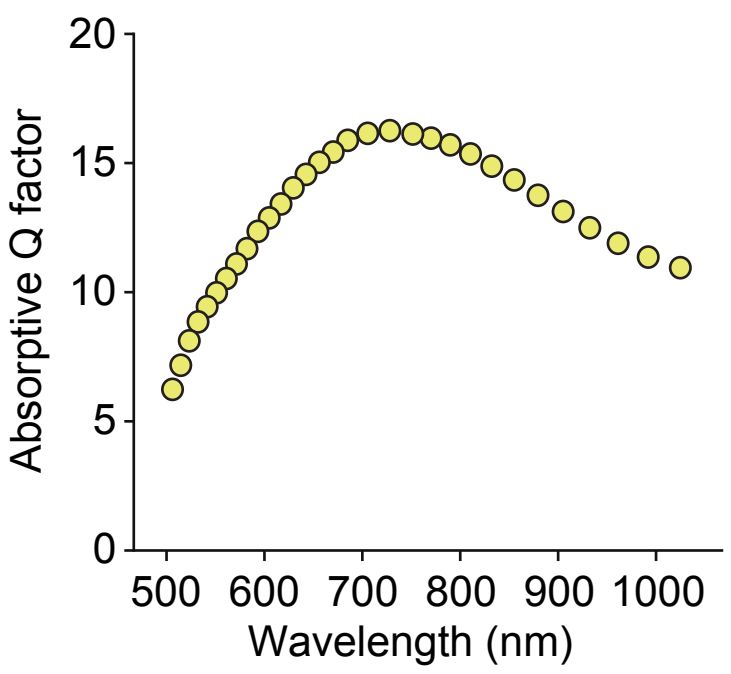

**Supplementary Fig. 4. Absorptive quality factor of gold**. Absorptive Q factor of the gold substrates calculated from complex refractive indices measured by ellipsometry as a function of wavelength.

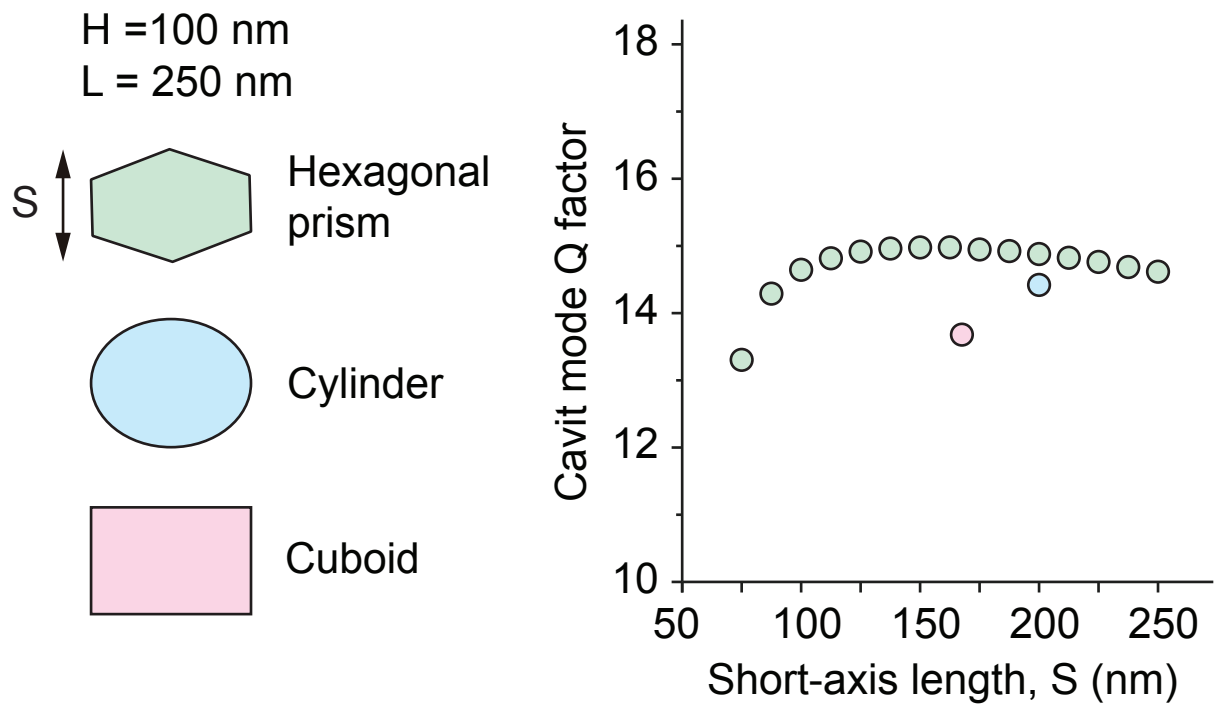

**Supplementary Fig. 5. Dependence of cavity Q factor on particle shape.** Simulated cavity quality factors of hexagonal-prism, cylindrical, and cuboidal particles with the same long-axis length of 250 nm and height of 100 nm, but different short-axis lengths, S. The cavity Q factor depends only weakly on particle shape.

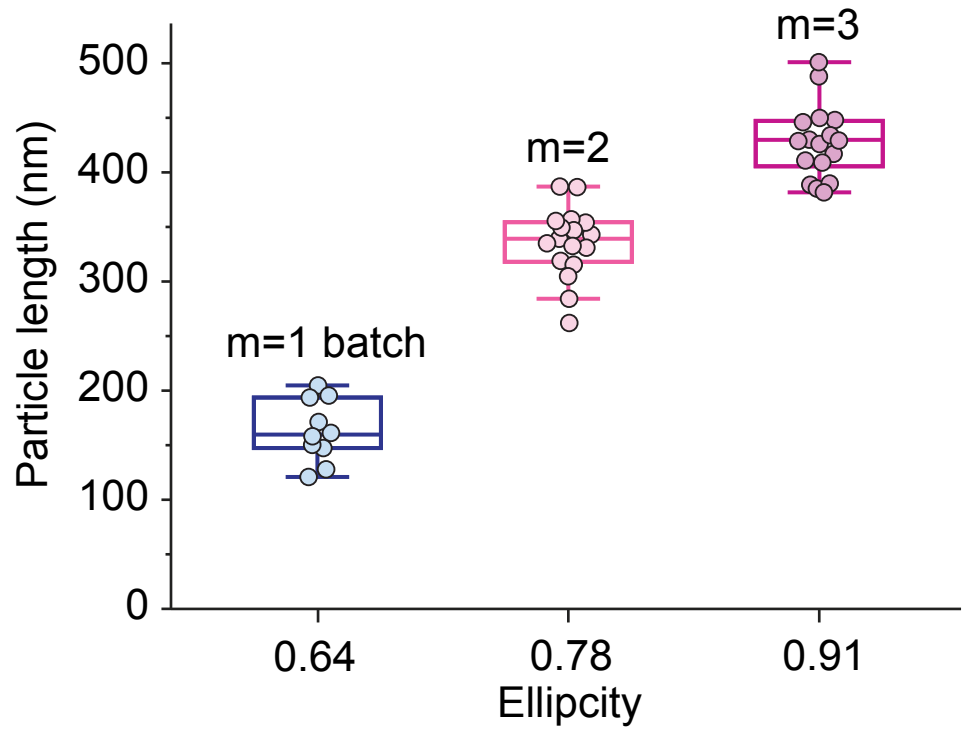

**Supplementary Fig. 6. Particle-size distributions and ellipticity of different fabrication batches.** Long-axis lengths of particles from different fabrication batches, measured from SEM images, shown as a function of ellipticity. The size ranges were selected to investigate the lowest-order m = 1 modes, second-order m = 2 modes, and third-order m = 3 modes, respectively. The distributions represent lateral particle sizes measured from SEM images. Because the m = 2 and m = 3 particles were fabricated in separate batches, differences in particle height and etch-induced ellipticity may contribute to distinct modal behavior despite partial overlap in lateral size.

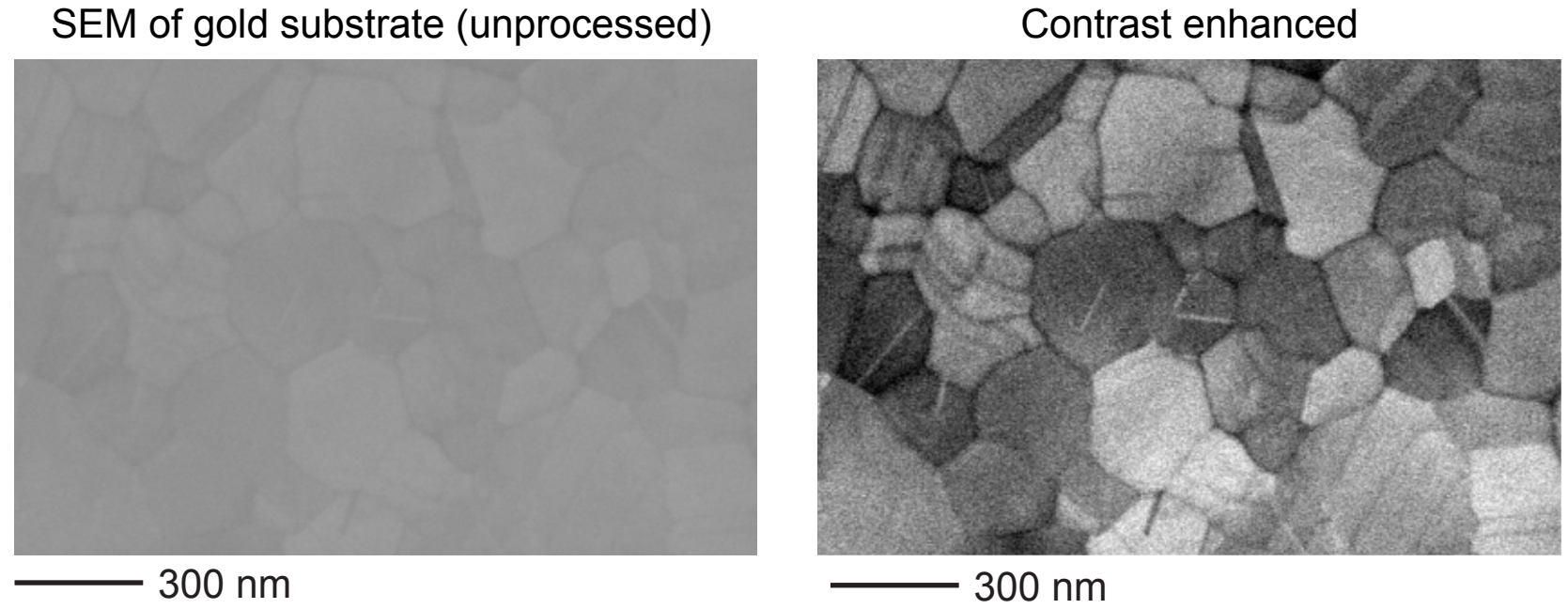

**Supplementary Fig. 7. Polycrystalline domains of the gold surface.** Representative SEM image of a gold substrate surface. The contrast-enhanced image, right, reveals polycrystalline domains with lateral sizes of approximately 150 to 450 nm.

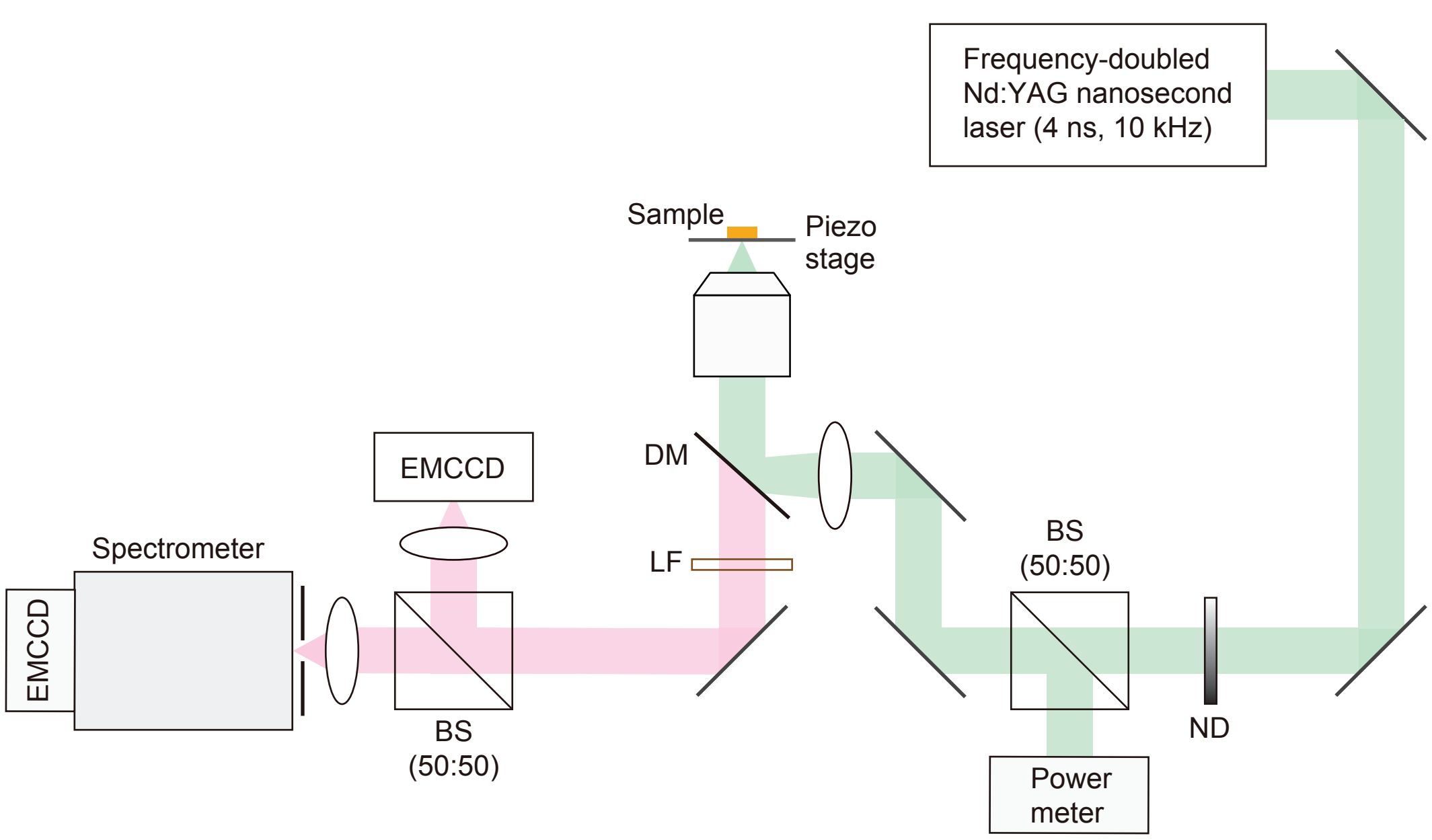

**Supplementary Fig. 8. Optical measurement setup.** ND, neutral density filter; BS, beam splitter; DM, dichroic mirror; LF, long-pass emission filter.

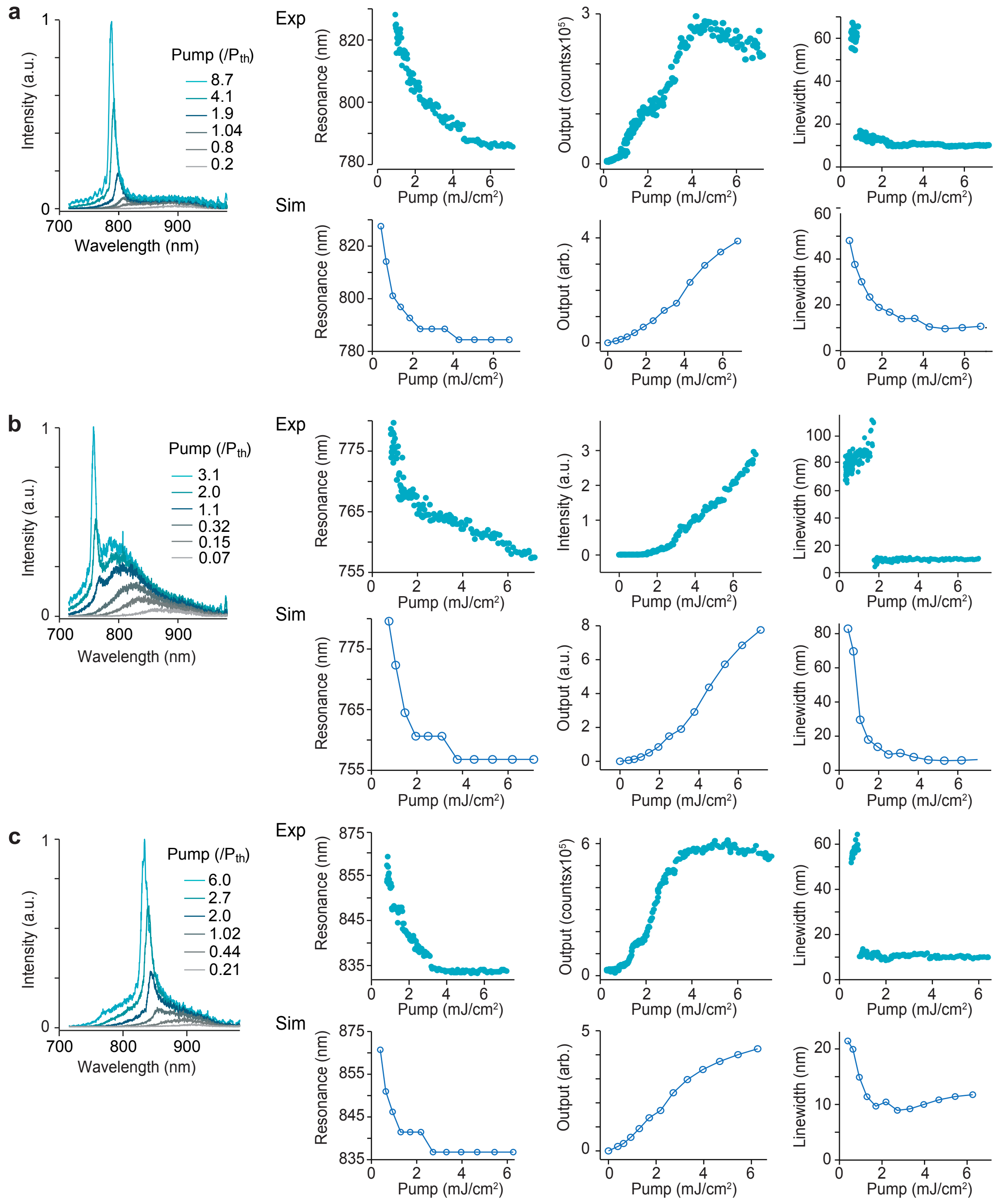

**Supplementary Fig. 9. Pump-dependent spectral characteristics of half-wave devices. a-c**, Three representative half-wave devices. Experimental results, top, and simulated results, bottom, are shown from left to right: resonance wavelength shift, light-in–light-out curves, and spectral linewidth as a function of pump fluence.